\documentclass[prb,twocolumn,showpacs,preprintnumbers]{revtex4}
\usepackage{graphicx}
\usepackage{subfigure}
\usepackage{amsfonts}
\usepackage{amssymb}
\usepackage{amsmath}
\usepackage{bm}

\begin{document}

\title{Direct and indirect excitons in semiconductor coupled quantum wells in an applied
electric field}

\author{K. Sivalertporn}
\author{L. Mouchliadis}
\altaffiliation{Present address: Department of Physics, University of Crete, 71003 Heraklion, Crete, Greece}
\author{A.\,L. Ivanov}
\author{R. Philp}
\author{E.\,A. Muljarov}
\altaffiliation[]{On leave from General Physics Institute RAS,
Moscow, Russia; egor.muljarov@astro.cf.ac.uk} \affiliation{School
of Physics and Astronomy, Cardiff University, The Parade, Cardiff
CF24 3AA, United Kingdom}

\date{\today}

\begin{abstract}

An accurate calculation of the exciton ground and excited states
in AlGaAs and InGaAs coupled quantum wells
(CQWs) in an external electric field is presented. An efficient and
straightforward algorithm of solving the Schr\"odinger equation in
real space has been developed and exciton binding energies,
oscillator strengths, lifetimes, and absorption spectra are
calculated for applied electric fields up to 100\,kV/cm. It is
found that in symmetric  8--4--8\,nm GaAs/Al$_{0.33}$Ga$_{0.67}$As CQW
structure,  the ground state of the system
switches from direct to indirect exciton at approximately 5\,kV/cm
with dramatic changes of its binding energy and oscillator
strength while the bright excited direct-exciton state remains
almost unaffected. It is shown that the excitonic lifetime is
dominated either by the radiative recombination or by tunneling
processes at small/large values of the electric field,
respectively. The calculated lifetime of the exciton ground state
as a function of the bias voltage is in a quantitative agreement
with low-temperature photoluminescence measurements. We have also
made freely available a numerical code for calculation of the
optical properties of direct and indirect excitons in
CQWs in an electric field.

\end{abstract}


\maketitle

\section{Introduction}

The electronic and optical properties of quantum well
structures have been widely investigated in the past decade due to
their potential applications in electro-optic and optoelectronic
devices.  In recent years there has been growing interest in
coupled quantum wells (CQWs) due to formation of
long-lived excitons when these structures are placed in an electric field (EF).
Intensive studies of indirect excitons in CQWs have resulted in their electrostatic and optical control.\cite{Butov02,Grosso09,Hammack09,Winbow11,Voros06,Voros09,Schinner11}
Very recently, CQWs have been embedded into Bragg-mirror microcavities and there has been found
a special type of voltage-tuned exciton polaritons which can be used
for optical nonlinearities and polariton lasing achieved at much lower threshold powers. \cite{Christmann11}

A CQW structure consists of two quantum wells separated by a barrier layer.
For a sufficiently thin barrier, the tunneling of carriers through the barrier makes
the two wells electronically coupled to each other.
As a result, an electron (hole) can either
reside in one of the two wells, or its wave function (WF) is distributed
between both wells. In the case of Coulomb bound electron and hole residing
in the same well, they form a direct exciton. If however they are located
in different wells, an indirect exciton is created.

In a symmetric CQW structure with no EF applied, formerly degenerate
single-particle states split, owing to the tunneling through the middle barrier,
into doublets with symmetric and antisymmetric
states in each. Since only transitions between states having the
same parity are optically allowed, the Coulomb-coupled electron-hole (e-h) pairs form excitonic states which are optically either
bright or dark. An EF, being applied in the growth direction, breaks down the symmetry of
the system making all these excitons bright. In fact, single-particle states experience with EF a transition
from states with well defined parity to the ones with the electron (hole) located in one of the two wells,
thus forming direct and indirect combinations of uncorrelated e-h pair states.
These different pair states are Coulomb coupled with each other and form an exciton
in which direct or indirect pair can dominate. In particular, with increasing EF, the
exciton ground state (GS) undergoes a transition from bright direct exciton to
indirect exciton which has a much weaker optical activity.
The exciton radiative lifetime increases due to a reduction
in the spatial overlap between the electron and hole WFs.
\cite{Polland85,Phillips89,Golub94,Alexandrou90,Charb88}
The exciton binding energy, in turn, reduces owing to an increased e-h
separation. It has also been found that the electronic coupling between quantum wells
considerably enhances the quantum-confined Stark effect in CQW structures.\cite{Chen87,Andrews88,Kim05,Le87}
The tunneling effect is also enhanced with the EF allowing the carriers to leak out of the system.\cite{Ahn86,Austin85}
This can lead to a considerable shortening of the photoluminescence decay time.\cite{Kash85}
All these properties of CQWs make them a much richer system compared to single quantum wells.

Excitonic states in CQWs in the presence of EF have been intensively studied in recent years.
Different theoretical approaches have been used ranging from variational methods\cite{Galbraith89,Kamizato89,Lee89,Dignam91,Linnerud94,Takahashi94,Soubusta99}
to direct diagonalizations in which the exciton WF is expanded into a large basis\cite{Szymanska03} or the Schr\"odinger equation is discretized in the momentum space.\cite{Arapan05}
In this paper, we present a more accurate and straightforward way for solving the Schr\"odinger equation for an exciton in a CQW structure. 
Expanding the exciton WF into e-h pair states we solve in the real space a system of differential
equations for the exciton in-plane motion, using the shooting method, here generalized to a matrix form.
The e-h pair basis states are calculated exactly using the analytical form of the electron (hole) WF in a uniform EF.
We present the full calculation of exciton bound and (discretized) continuum states as well as the absorption spectrum of a CQW in an applied EF. We study the EF effect on the exciton binding energy and lifetime in 8-4-8\,nm
GaAs/Al$_{0.33}$Ga$_{0.67}$As symmetric CQW structure, both for the ground and excited states, and
demonstrate a direct-to-indirect crossover of the exciton ground state with increasing EF.
An example of an asymmetric 10-4-10\,nm InGaAs CQW having different In content in the left and right QWs is also given, demonstrating our calculation of the electron and hole energies of quantization and the exciton transition energies and oscillator strengths as functions of the applied EF.

\section{Formalism and numerical method}

Let us consider a symmetric GaAs/Al$_{x}$Ga$_{1-x}$As CQW which
consists of two GaAs QW layers separated by a thin Al$_{x}$Ga$_{1-x}$As barrier
and surrounded on both sides by thick barriers of the same kind. In this paper we mainly concentrate on a
CQW structure which has been intensively used in a series of experiments,\cite{Butov02,Grosso09,Hammack09,Winbow11,Butov99} taking the barrier and well widths to be $L_b=4$\,nm and $L_w=8$\,nm, respectively, and the barrier
concentration of Al to be $x=0.33$. The electric field $F$ is applied in the growth direction.
For a different CQW in presence of an EF simulations can be performed using our online available numerical code\cite{OnlineCode}
(see Sec.\,III\,D below for more details).
We are interested in optically allowed transitions in such a system and
thus consider excitonic states with zero in-plane and angular
momenta only. In the effective mass approximation, the excitonic
Hamiltonian can be divided into three parts: the first two,
$\hat{H}_{e}$ and $\hat{H}_{h}$, take into account the electron
and hole quantization in heterostructure potentials $V_e$ and
$V_h$ and the third one, $\hat{H}_{X}$, is responsible for the
electron-hole (e-h) in-plane relative motion and Coulomb binding:
\begin{equation}
\hat{H}(z_e,z_h,\rho) = \hat{H}_{e}(z_e) + \hat{H}_{h}(z_h) +
\hat{H}_{X}(z_e,z_h,\rho) +E_g\label{Ham_full}
\end{equation}
with
\begin{eqnarray}
\hat{H}_{e,h}(z)= -\frac{\hbar^{2}}{2}\frac{\partial}{\partial
z}\frac{1}{m_{e,h}(z)}\frac{\partial}{\partial z} + V_{e,h}(z) \pm
e F z,
 \label{H_eh}
\\
\hat{H}_{X}=-\frac{\hbar^{2}}{2 \mu}
\left(\frac{\partial^{2}}{\partial \rho^{2}} + \frac{1}{\rho}
\frac{\partial}{\partial \rho}\right) -
\frac{e^{2}}{\varepsilon_{b} \sqrt{(z_{e}-z_{h})^{2}+
\rho^{2}}}\,,
 \label{H_X}
\end{eqnarray}
where $z_{e(h)}$ is the electron (hole) coordinate in the growth
direction, $\rho$ the coordinate of the e-h relative
motion in the QW plane, $\varepsilon_{b}$ the background
dielectric constant which we assume to be $z$-independent, $m_{e}$
the electron effective mass, and $E_g$ the bandgap of the well
material (GaAs). Owing to the strong QW confinement, the
heavy-hole subband is split off considerably and can be
approximated by an anisotropic effective mass using the
Kohn-Luttinger parameters $\gamma_{1}$ and $\gamma_{2}$:\cite{KL}
\begin{eqnarray}
\frac{1}{m_h} &=& \frac{1}{m_{0}}(\gamma_{1} - 2\gamma_{2})\,, \\
\frac{1}{\mu} &=& \frac{1}{m_{e}} + \frac{1}{m_{0}}(\gamma_{1} +
\gamma_{2})\,,
\end{eqnarray}
where $m_h$ is the hole effective mass in the growth direction,
$\mu$ the exciton in-plane reduced mass, and $m_{0}$ the
free electron mass. We assume a rectangular form of the
heterostructure confinement potentials,
\begin{equation}
V_{e,h}(z)= \left\{
\begin{array}{ll}
       \ 0 & \rm{\ inside~the~wells},  \\
       {\cal{V}}_{e,h}>0 & \rm{\ outside},
\end{array}
\right.
 \label{Poten_z}
\end{equation}
and similar step-like profiles for the electron and hole effective masses in
the growth direction.

In our calculation, we have used the following parameters: the
background dielectric constant\cite{Dignam91} $\varepsilon_{b}=12.5$, the
energy-band offset ratio\cite{Arapan05}
${\cal{V}}_{e}\!:\!{\cal{V}}_{h}=65\!:\!35$. The band gap
discontinuity at the GaAs/Al$_{x}$Ga$_{1-x}$As interface is
linearly approximated\cite{Dignam91} as $1.247\cdot x$\,eV. The
Kohn-Luttinger parameters for pure GaAs and AlAs are obtained from
Ref.\,\onlinecite{Molenkamp88}: $m_{e}=0.0665\,m_{0}$,
$\gamma_{1}=6.79$ and $\gamma_{2}=1.92$ in GaAs (giving the hole mass
$m_{h}=0.34\,m_{0}$); $m_{e}=0.15\,m_{0}$,
$\gamma_{1}=3.79$ and $\gamma_{2}=1.23$ in AlAs (giving
$m_{h}=0.75\,m_{0}$). The Al$_{x}$Ga$_{1-x}$As alloy
parameters were linearly interpolated between those of GaAs and
AlAs. In particular, for the content $x=0.33$, we have used
$m_{e}=0.094\,m_{0}$ and $m_{h}=0.48\,m_{0}$ in
Al$_{0.33}$Ga$_{0.67}$As layers. The in-plane reduced mass in the barrier layers ($0.057\,m_{0}$) is different
from that in the well layers ($0.042\,m_{0}$). However, because of a very small probability for the exciton ground state
to find the carriers in the barrier, we take for the in-plane reduced mass the GaAs value
of $\mu=0.042\,m_{0}$.

\subsection{Electron and hole single-particle states}
\label{Single}

To solve the excitonic Schr\"odinger equation with the full
Hamiltonian Eq.\,(\ref{Ham_full}) we first prepare a basis of
single-particle states which satisfy the following one-dimensional
equations:
\begin{equation}
\hat{H}_{e,h} (z) \psi^{e,h}(z) = E^{e,h}\psi^{e,h}(z)\,.
 \label{SEz}
\end{equation}
To do so, we use the advantage of the analytic form of the
electron and hole WFs in the rectangular confinement potentials
Eq.\,(\ref{Poten_z}) and uniform electric field $F$. In each layer
of the CQW structure, the electron WF is given by a superposition
of two Airy functions\cite{Abramowitz}
\begin{equation}
\psi^{e}(z)=  a_k{\rm Ai}(\xi)+ b_k{\rm Bi}(\xi)\,,
 \label{psi_e}
\end{equation}
where
\begin{equation}
\xi(z) = \left(\frac{2 m_{e} eF}{\hbar^{2}}\right)^{1/3} \left[z -
\frac{E^e - V_e(z)}{eF}\right]
\end{equation}
and the index $k$ labels the heterostructure layers (from left to
right) taking integer values from 1 to 5. The electron
eigenenergy $E^e$ and five pairs of coefficients ($a_k$,\,$b_k$)
in Eq.\,(\ref{psi_e}) are found from four pairs of boundary
conditions (BCs) on heterostructure interfaces and two BCs at
$z\to\pm\infty$.

The interface BCs following from Eqs.\,(\ref{H_eh}) and
(\ref{SEz}) are the continuity of $\psi^e(z)$ and $m_e^{-1}(z)
\partial \psi^e(z)/\partial z$. The other two BCs take into account
the possibility for the electron to tunnel through the barrier and
escape from the system to the side of the CQW structure where the
applied EF gradually lowers the potential. In that area,
the solution is given by a wave
propagating away from the system. For the electron WF and $F>0$, this outgoing BC
at $z\to-\infty$ yields $b_1=-ia_1$ which follows
from the specific combination of the Airy functions ${\rm
Ai}(\xi)-i{\rm Bi}(\xi)$ producing an outgoing
wave.\cite{Ahn86,Abramowitz} At the same time, the electron cannot escape
to the other side of the structure where the potential
gradually increases, and thus the other BC, typical for bound/localized states,
is $\psi^e(z\to+\infty)=0$, giving
$b_5=0$ due to the asymptotics of the Airy functions.\cite{Abramowitz}
The form of the WF and the BCs for the hole are found in a similar
way, taking into account that the potential grows in the opposite direction.
The secular equation following from all ten BCs
determines discrete eigenvalues of Eq.(\ref{SEz}),
\begin{equation}
E_j^{e,h} = \tilde{E}_j^{e,h} - i \Gamma_j^{e,h}\,,
 \label{E-eh}
\end{equation}
which are the complex energies of electron/hole resonant states, also known in
the literature as Siegert states.\cite{Siegert39} The real part
of the eigenvalue, $\tilde{E}_j^{e(h)}$, is the energy position of
the electron (hole) $j$-th resonant level, while the imaginary
part $\Gamma_j^{e(h)}$ gives its tunneling linewidth.

The WF of any resonant state having a finite linewidth is
essentially complex, i.\,e. it cannot be made real by any uniform
phase shift. Also, its amplitude grows exponentially to the outside area to which the
particle can escape and thus has to be normalized to its
flux.\cite{Siegert39,More71} This normalization includes a divergent volume
integral and a compensating surface term. For the values of the EF
considered in this paper, the calculated linewidths
of the electron and hole states of interest are always small compared to their
energies of quantization. Similarly, the imaginary parts of the WFs are
small compared to the real ones and can be dropped. The
normalization condition is then taken in a form
\begin{equation}
\int^{z_{\rm max}}_{z_{\rm min}}
\bigl[\tilde{\psi}^{e,h}_j(z)\bigr]^{2} dz =1\,,
 \label{Norm}
\end{equation}
where $\tilde{\psi}^{e,h}_j={\rm Re} (\psi^{e,h}_j)$. The limits of integration in Eq.\,(\ref{Norm}),
$z_{\rm min}$ and $z_{\rm max}$, taken to be the same for electron and hole, are two distant points on
both sides of the CQW where the WFs decay considerably before they start to grow exponentially
owing to the carrier tunneling, so that the
surface terms are minimized and can be dropped leaving in the normalization only finite-volume
integrals.

  \begin{figure}[t]
  \begin{center}
\includegraphics[width=\columnwidth]{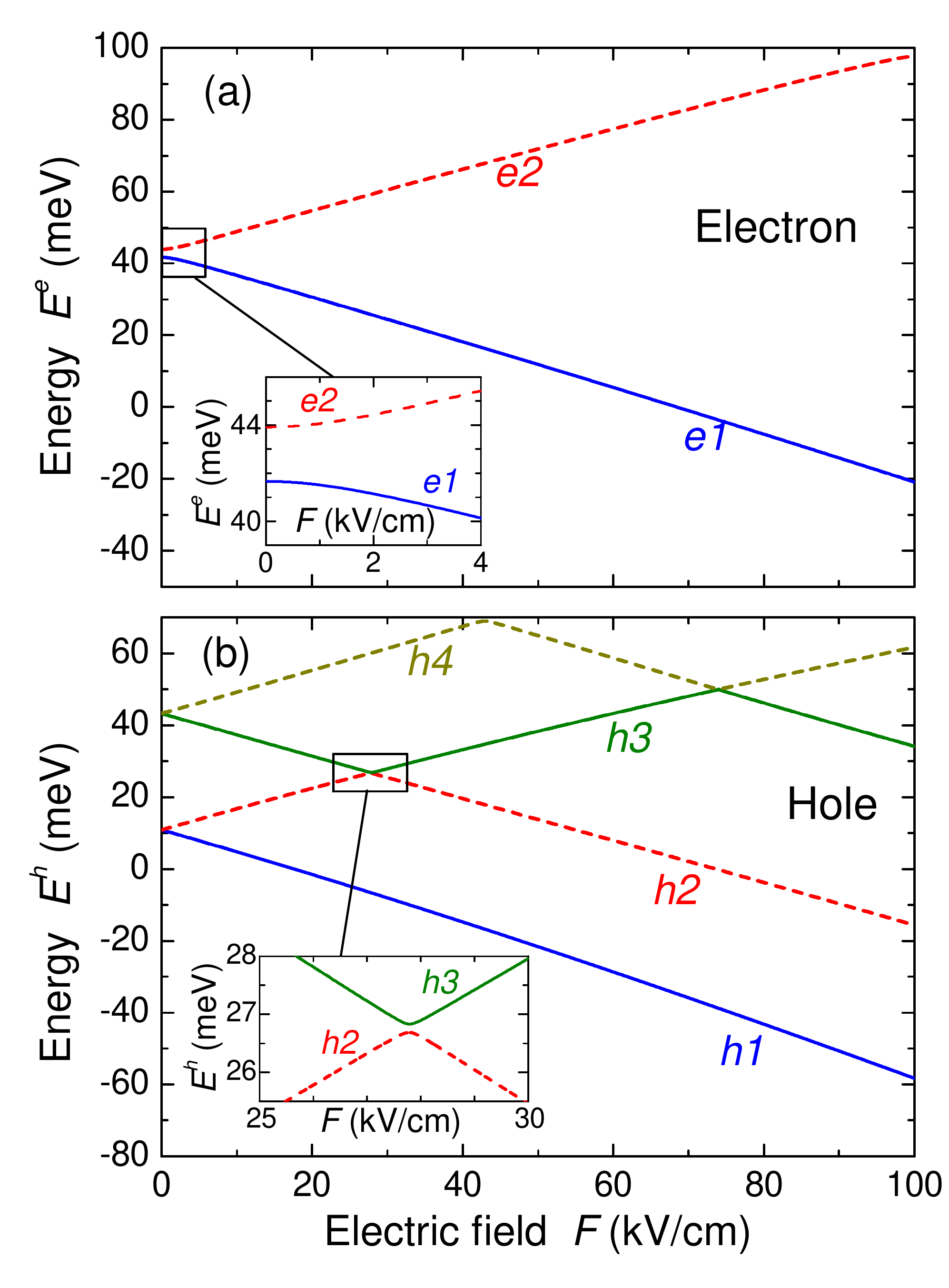}
  \caption{Energies of two electron states (a) and four hole states (b) in 8--4--8\,nm GaAs/Al$_{0.33}$Ga$_{0.67}$As CQW as functions of an applied electric field. Insets zoom in particular spectral regions with anticrossing.
\label{fig:Energy}}
  \end{center}
  \end{figure}

  \begin{figure}[t]
  \begin{center}
 \includegraphics[width=\columnwidth]{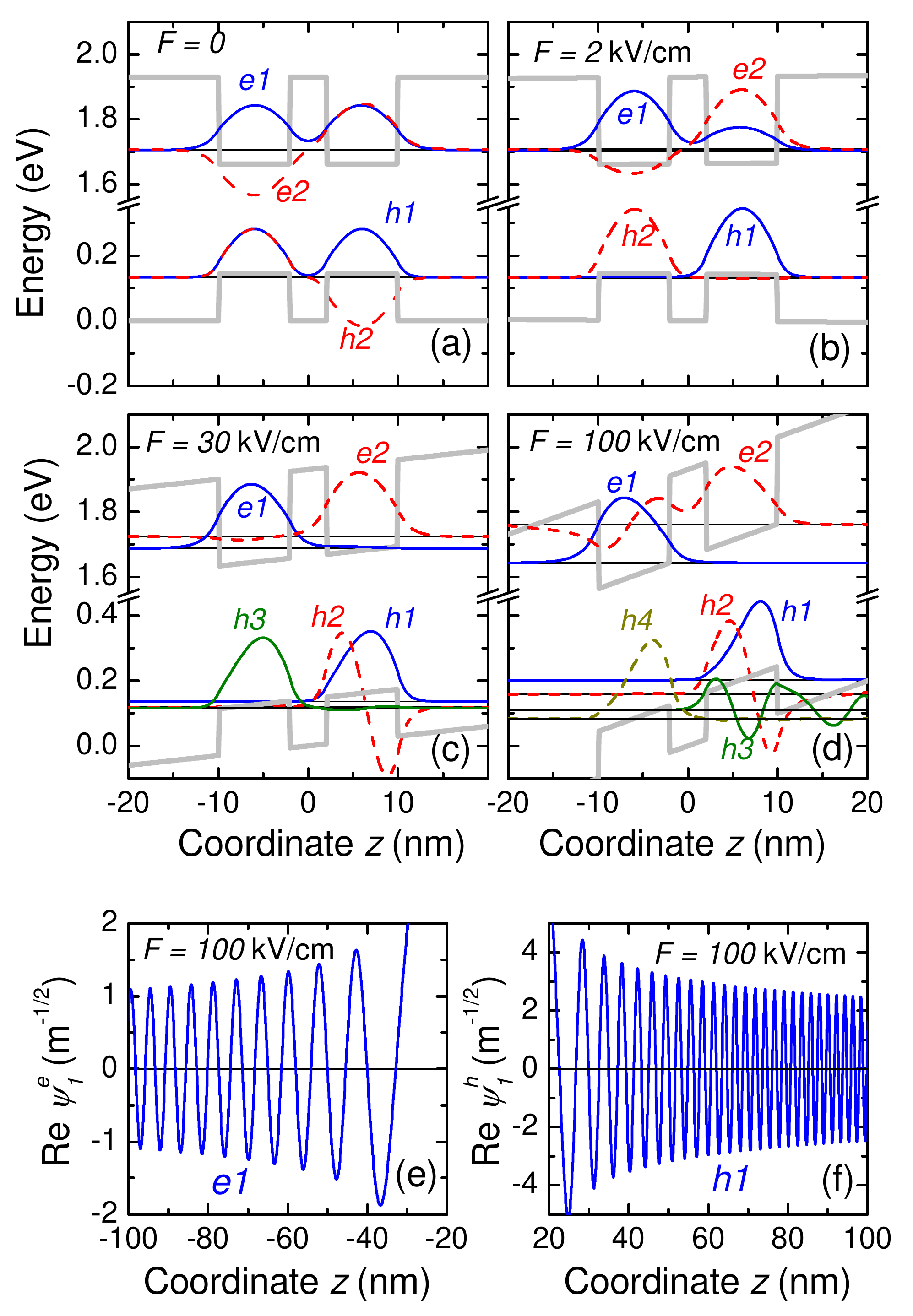}
  \caption{(a)-(d) Wave functions and energy levels of electron and hole ground and excited states
for different values of the electric field $F$. Grey lines show CQW heterostructure potentials, using $E_{g}=1.519$\,eV
for the band gap. (e),(f) Oscillating tails in the electron and hole wave functions.
\label{fig:psiz}}
  \end{center}
  \end{figure}

Different electron and hole subbands ($\tilde{E}^{e}_{i}$ for $i=1,2$ and
$\tilde{E}^{h}_{j}$ for $j=1$\,-\,4) calculated in the presence of the EF
are shown in Fig.\,\ref{fig:Energy}.  The corresponding WFs, $\tilde{\psi}^e_i$
and $\tilde{\psi}^h_j$, are illustrated in
Fig.\,\ref{fig:psiz}(a)-(d) for a few different values of the EF.
At zero field, the GS and the first ES have,
respectively, symmetric and antisymmetric WFs, see Fig.\,\ref{fig:psiz}(a).
With increasing EF, the WFs become asymmetric, and the WF maxima for the electron and
hole GSs move in opposite directions. The GS and the first ES for the same carrier are also
confined in different QWs. This happens to both carriers already at $F=2$\,kV/cm, see Fig.\,\ref{fig:psiz}(b). As the EF grows further, the GS-ES splittings increase almost linearly with $F$ (Fig.\,\ref{fig:Energy}), and the hole
ES jumps from the left to the right QW [Fig.\,\ref{fig:psiz}(c)]. This corresponds to an
anticrossing of hole ES subbands which takes place at 27.8\,kV/cm,
see the inset in Fig.\,\ref{fig:Energy}(b). Such an anticrossing
behavior was also found in previous calculations.\cite{Lee89,Arapan05} The
same happens to the electron ES at a much higher EF.
Figure~\,\ref{fig:psiz}(d) shows a modified ES WF for
$F=100$\,kV/cm which is a precursor of a similar transition for
the electron.

Far from the CQW structure, we observe tiny oscillations in the WF of the
electron (hole) GS in the left (right) barrier, see
Fig.\,\ref{fig:psiz}(e,f). This oscillatory behavior, typical for
freely propagating particles,
occurs due to the lowering of the potential by the EF, so that an
electron (hole) can escape to the left (right) barrier. Since
$m_h V_{h}> m_e V_{e}$, the frequency of the oscillations for the
hole is larger than that for the electron. At the same time, the
amplitude of oscillations is also larger for the hole since
the hole tunneling is stronger due to $V_{h}< V_{e}$.
Such oscillations become more dramatic and start earlier (closer to the CQW center)
for some of higher ESs. For example, for $F=100$\,kV/cm  the hole state $h3$ exibits huge oscillations
clearly seen in Fig.\,\ref{fig:psiz}(d). However, the next hole ES $h4$ which has the dominant contribution to the direct
exciton state discussed below has only tiny oscillations, similar to those in Fig.\,\ref{fig:psiz}(f), and a small tunneling rate, comparable to that of the hole GS.

\subsection{Excitonic states: Multi-sublevel approach}

We calculate excitonic states in a CQW structure in the presence
of an EF, expanding the exciton WF into a finite set of
e-h pair states:
\begin{equation}
\Psi (z_{e},z_{h}, \rho) = \sum^{N}_{n=1} \Phi_n (z_{e}, z_{h})
\phi_n (\rho)\,,
 \label{ex-wf}
\end{equation}
where
\begin{equation}
\Phi_n (z_{e}, z_{h})=\tilde{\psi}^{e}_{i} (z_{e})
\tilde{\psi}^{h}_{j} (z_{h})\,,\ \ \ n=(i,j)\,,
 \label{pair}
\end{equation}
and $\tilde{\psi}^{e,h}_{i} (z)$ are the electron and hole wave
functions calculated in the presence of EF and heterostructure
potentials, see Sec.\,\ref{Single}. The Schr\"odinger equation for
the exciton, $\hat{H}\Psi = E_X\Psi$, then takes the form
\begin{equation}
\left[ \hat{K}({\rho}) +E^{(0)}_n-E_X
\right] \phi_n(\rho)+
\sum^{N}_{m=1} V_{nm}(\rho)\phi_m(\rho) =0
 \label{matrix-se}
\end{equation}
with
\begin{equation}
\hat{K}({\rho}) = -\frac{\hbar^{2}}{2 \mu}
\left(\frac{\partial^{2}}{\partial \rho^{2}} + \frac{1}{\rho}
\frac{\partial}{\partial \rho} \right)\,,
 \label{kinetic}
\end{equation}
\begin{equation}
 V_{nm} (\rho)=
-\frac{e^{2}}{\varepsilon_{b}}
\int\!\!\!
\int^{z_{\rm max}}_{z_{\rm min}} \frac{\Phi_n (z_{e}, z_{h})\Phi_m (z_{e},
z_{h})}{\sqrt{(z_{e}-z_{h})^{2}+ \rho^{2}}}\, dz_{e} dz_{h} \,,
\end{equation}
and the energies of pair states given by
\begin{equation}
E^{(0)}_n=\tilde{E}^e_i+\tilde{E}^h_j+E_g\,.
\label{E0n}
\end{equation}

In our calculation of the exciton states in 8-4-8\,nm
GaAs/Al$_{0.33}$Ga$_{0.67}$As CQW for EFs up to $F=25$\,kV/cm
it was sufficient to restrict the basis in Eq.\,(\ref{ex-wf}) to
four e-h pair states ($N=4$), keeping only the GS and
the first ES for electron and hole. We
label these four basis states as $e1h1$ ($n=1$), $e1h2$ ($n=2$), $e2h1$ ($n=3$), and
$e2h2$ ($n=4$), where $e$\,($h$) stands for an electron (hole) and the
numbers 1 and 2 refer to the single-particle GS and first ES,
respectively, while $n$ is the unified pair index introduced in Eq.\,(\ref{pair}).
Other QW structures may require higher ESs to be taken into account.
These are also needed in our case when a stronger EF is considered.
In particular, higher quantized levels for the hole $h3$ and $h4$ [see Fig.\,1(b)]
are taken into account for $F>25$\,kV/cm and $F>70$\,kV/cm, respectively, and these states
have a major contribution to the direct exciton WF.

We calculate the exciton transition energy $E_X$ and the in-plane
components of the WF, $\phi_n(\rho)$, by solving the
matrix differential equation (\ref{matrix-se}) numerically. To do so, we
introduce a matrix generalization of the shooting method applying
the latter to a system of coupled differential equations. The
shooting method transforms a boundary-value problem like
Schr\"odinger's equation with BCs to an initial-value problem in which one
of the boundary values (in the present case the WF at
$\rho\to\infty$) is taken as a starting point. The boundary
value(s) on the other side (at $\rho=0$) is then used to find the
eigenenergies. The BCs follow straightforwardly from
Eq.\,(\ref{matrix-se}) and the asymptotics
of the Coulomb matrix elements $V_{nm}(\rho)$. At large distances
$V_{nm}(\rho)\to-\delta_{nm} e^2/(\varepsilon_b \rho)$, while
at small distances the potentials $V_{nm}(\rho)$ have
logarithmic dependence, as is clear from Fig.\,3. Therefore for bound
states
\begin{equation}
\phi_n(\rho\to \infty)=A_n \rho^{s_n} e^{-\alpha_n\rho}\,,
 \label{phi-inf}
\end{equation}
where $\alpha_n=\sqrt{2\mu(E_n^{(0)}-E_X)}/\hbar$ and
$s_n=\mu e^2/(\hbar^2\varepsilon_b\alpha_n)-1/2$,
and
\begin{equation}
\phi_n'(0)=0\,.
 \label{bc0}
\end{equation}

  \begin{figure}[t]
  \begin{center}
\includegraphics[width=\columnwidth]{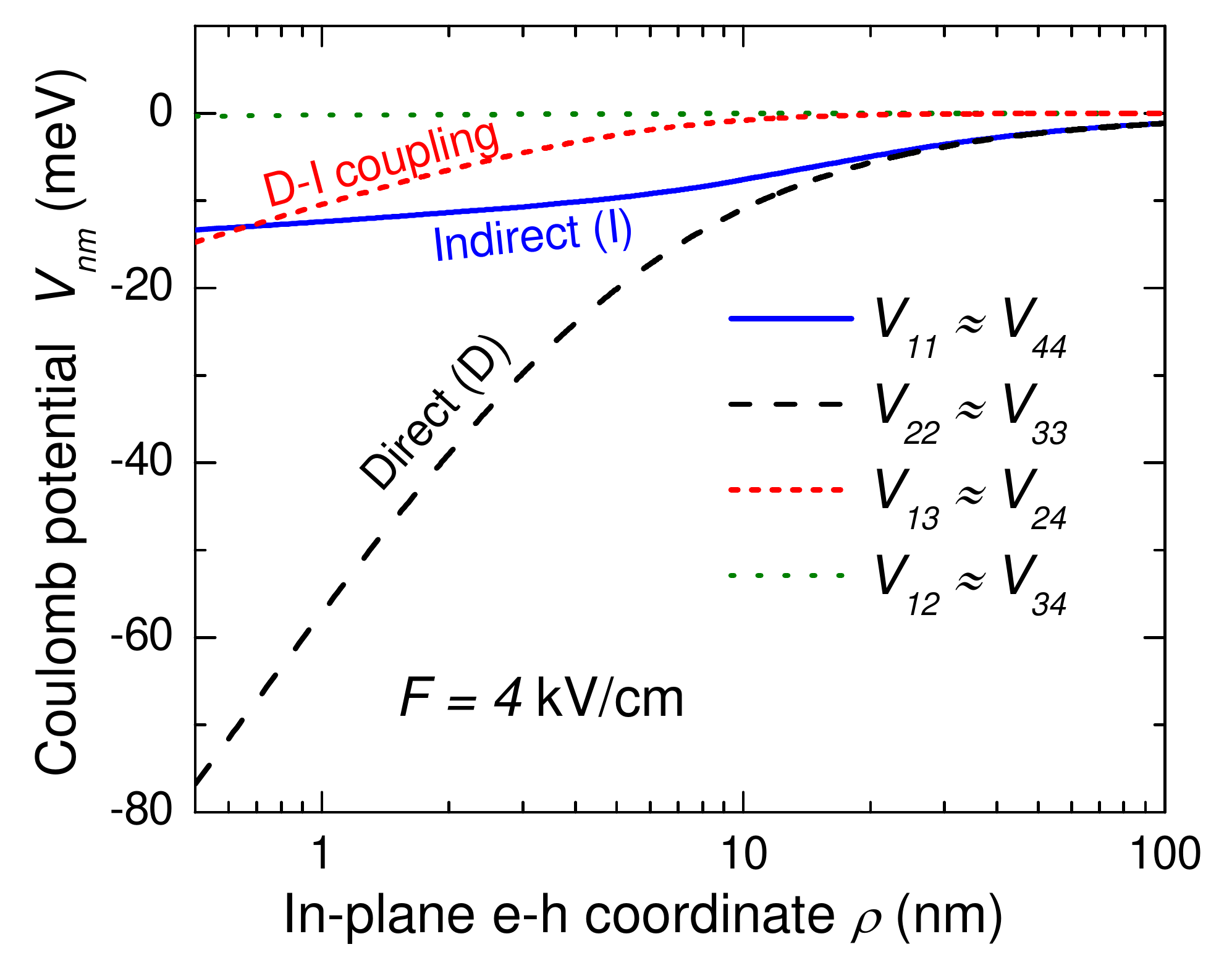}
  \caption{Matrix elements $V_{nm}$ of the Coulomb potential calculated between different e-h pair states at $F$=4\,kV/cm.
   \label{fig:Veh}}
  \end{center}
  \end{figure}
Discretizing Eq.\,(\ref{matrix-se}) on a finite grid, a numerical
solution in the area $0\leq\rho\leq R$ is generated iteratively
using a finite difference scheme. In particular, a second-order
scheme which we have used in our calculation brings
Eq.\,(\ref{matrix-se}) to the form
\begin{equation}
\phi_n(\rho-\Delta\rho)=-\phi_n(\rho+\Delta\rho)\frac{2\rho+\Delta\rho}{2\rho-\Delta\rho}+\sum_{m=1}^N
F_{nm}(\rho)\phi_m(\rho)\,,
\end{equation}
where $\Delta\rho$ is the discretization step and the matrix
$F_{nm}(\rho)$ depends on the Coulomb interaction and on a
finite-difference representation of the kinetic term
Eq.\,(\ref{kinetic}). The WF amplitudes $A_n$ in the
starting values $\phi_n(R)$ in Eq.\,(\ref{phi-inf}) are the unknowns
to be found along with the eigenvalue $E_X$. Going to the very last
point $\rho=0$ and using the boundary condition
Eq.\,(\ref{bc0}) produces a homogeneous matrix equation of the form
\begin{equation}
\sum_{m=1}^N M_{nm}(E_X) A_m=0
 \label{me}
\end{equation}
in which $M_{nm}(E_X)$ depends solely on the exciton energy $E_X$ (and
not on $\rho$ any more), and thus the energy eigenvalues are
determined by
\begin{equation}
\det|M_{nm}(E_X)|=0\,.
\end{equation}

For small values of the EF, the electron and hole GS-ES splittings
are smaller than the exciton Coulomb energy (compare Figs.\,1 and 3)
and thus several bound states [having the asymptotics given by
Eq.\,(\ref{phi-inf})] can always be found in the system. However,
as the EF grows, the Coulomb energy of the exciton ESs
is getting smaller than the e-h pair splitting energies and thus
some of these exciton states become unbound. Since the unbound states
have energies $E_X>E_1^{(0)}$, at least for some of their radial
components the asymptotics Eq.\,(\ref{phi-inf}) is no longer valid
and a proper treatment of the excitonic continuum is required.
This task is outside the scope of the present paper which mainly
concentrates on exciton bound states. Nevertheless, some effects of the
continuum and in particular its influence on the excitonic
absorption spectrum can be taken into account, in a first attempt, by
restricting the exciton in-plane motion to a large circle of
radius $R$ and in this way discretizing the continuum. The
asymptotic boundary conditions Eq.\,(\ref{phi-inf}) are now replaced
by
\begin{equation}
\phi_n(R)=0\,,\ \ \ \phi_n(R-\Delta\rho)=A_n\,,
\label{circle}
\end{equation}
where the new amplitudes $A_n$ satisfy the same Eq.\,(\ref{me})
with matrix $M_{nm}(E)$ being redefined accordingly.

It is convenient to normalize the radial components of the WFs
introducing expansion coefficients $C_n$:
\begin{equation}
\phi_n(\rho)=C_n\tilde{\phi}_n(\rho)\,,
 \label{C-intro}
\end{equation}
where $\tilde{\phi}_n(\rho)$ is normalized to $2\pi\int_0^\infty|\tilde{\phi}_n|^2 \rho d \rho=1$, and therefore
\begin{equation}
\sum_{n=1}^N|C_n|^2=1\,,
\end{equation}
due to orthogonality of the e-h pair states $\Phi_n (z_{e},
z_{h})$ and normalization of the total exciton WF $\Psi
(z_{e},z_{h}, \rho)$.

Finally, for each
excitonic state, the oscillator strength per unit area is
calculated as\cite{Andreani95}
\begin{equation}
f =\frac{2m_0 E_X \left|d_{cv}\right|^{2}}{\hbar^2} \left|\int^{z_{\rm
max}}_{z_{\rm min}}\Psi (z,z,\rho=0)\, dz\right|^{2}\,,
 \label{Osc}
\end{equation}
where $d_{cv}$ is the basic dipole matrix element between the valence and
conduction bands, and the overlap integral in
Eq.\,(\ref{Osc}) accounts for the spatial distribution of the excitonic recombination. The exciton
radiative linewidth is then given by
\begin{equation}
\Gamma_{R} = \frac{\pi e^{2}\hbar}{\sqrt{\varepsilon_b} m_{0} c}
f\,,
 \label{Rad}
\end{equation}
where $c$ is the speed of light.

\section{Results and discussion}

In the presence of EF, the energy levels of the electron and hole single-particle states in a
CQW structure experience
Stark shifts $\tilde{E}^{e (h)}_{1,2}(F)\approx\tilde{E}^{e (h)}_{1,2}(0)\mp dF/2$ as
demonstrated in Fig.\,1, where $d=L_w+L_b$ is the center-to-center distance between the QWs.
Concentrating on these two lowest
levels for the electron and two for the hole we are thus dealing with four
e-h pair states. For two of them, $e1h2$ and $e2h1$, the energies $E^{(0)}_{2,3}$ [see Eq.\,(\ref{E0n})]
remain almost unaffected by the
EF as the Stark shift for the electron is compensated by that for the hole, while the other
two pair states, $e1h1$ and $e2h2$, have twice larger Stark shifts than the
single-particle states:
$E^{(0)}_{1,4}(F)\approx E^{(0)}_{1,4}(0)\mp dF$.
Even though our CQW is symmetric, at
nonzero EF  the ground and the first excited single-particle states
are localized in different wells of the CQW structure, see Fig.\,2(b). That is why the pair
states with electron and hole in the same QW ($n=2$  and $n=3$) are electrically neutral
and can be called {\em direct} states while the other two, $n=1$ and $n=4$, with electron and hole in
different QWs have nonzero dipole moment and are called {\em indirect} states.
As for excitonic states, they are, strictly speaking, neither direct nor indirect, since the exciton WF is
always a combination of different pair states.

\subsection{Crossover from direct to indirect exciton}

In the absence of the Coulomb interaction, all four pair states are nearly degenerate at zero EF: Due to a rather weak
tunneling through the middle barrier of the CQW,
the energy splittings are small compared to the Coulomb energy (2.3\,meV and 0.05\,meV splitting for electron and
hole, respectively, versus 4--9\,meV binding energy). The Coulomb interaction splits further and strongly mixes the direct and indirect e-h pair states. To understand these splitting and mixing, let us consider the Coulomb matrix elements in more detail.

  \begin{figure}[t]
  \begin{center}
\includegraphics[width=\columnwidth]{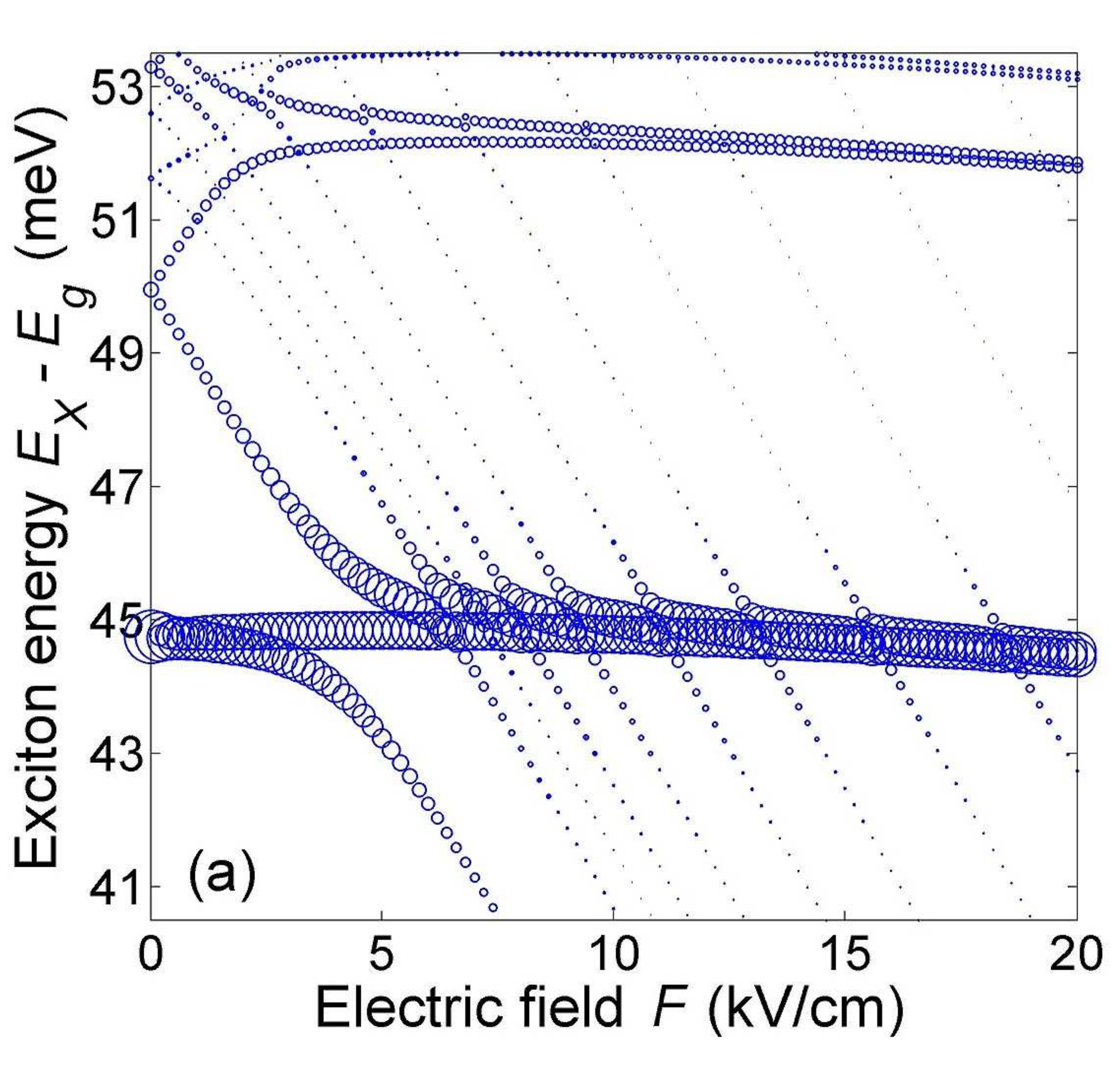}
\vskip5mm
\includegraphics[width=\columnwidth]{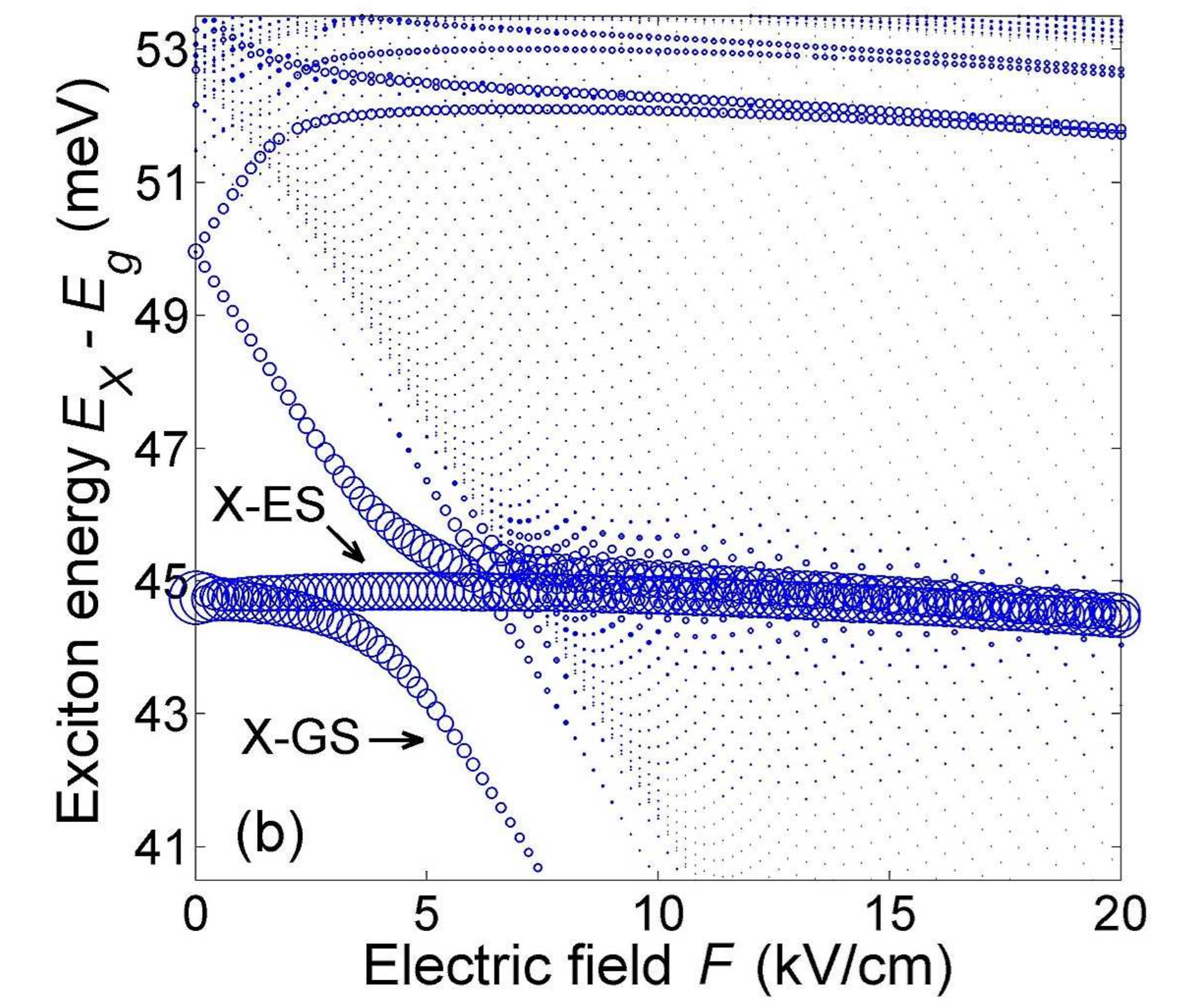}
  \caption{ Electric field dependence of the optical transition energy $E_X$ for different exciton states in 8-4-8\,nm
GaAs/Al$_{0.33}$Ga$_{0.67}$As symmetric CQW structure, calculated using the exciton confinement radius $R=200$\,nm (a) and 800\,nm (b). The circle area is proportional to the exciton oscillator strength $f$.
\label{fig:osc}}
  \end{center}
  \end{figure}

Figure~3 shows an example of the Coulomb matrix elements for $F=4$\,kV/cm, though this picture does not change much when the EF increases or decreases. The direct and indirect pair states have different charge separation, and thus their diagonal Coulomb matrix elements are also quite different: Potentials $V_{22}\approx V_{33}$ for the direct pairs are a few times stronger than $V_{11}\approx V_{44}$ for the indirect ones. This difference brings in a considerable splitting between the direct and indirect pair states at $F=0$: The indirect doublet is found almost 5\,meV above the direct one, see Fig.\,4. Among off-diagonal elements, the largest are $V_{13}\approx V_{24}$, due to a considerable overlap between the electron ground and excited states. All other matrix elements are two-three orders of magnitude smaller, because of a much larger effective mass of the hole and consequently much smaller overlap integrals. All off-diagonal elements drop quickly at large $\rho$ due to the orthogonality of WFs.
 \begin{figure}[t]
 \begin{center}
\includegraphics[width=\columnwidth]{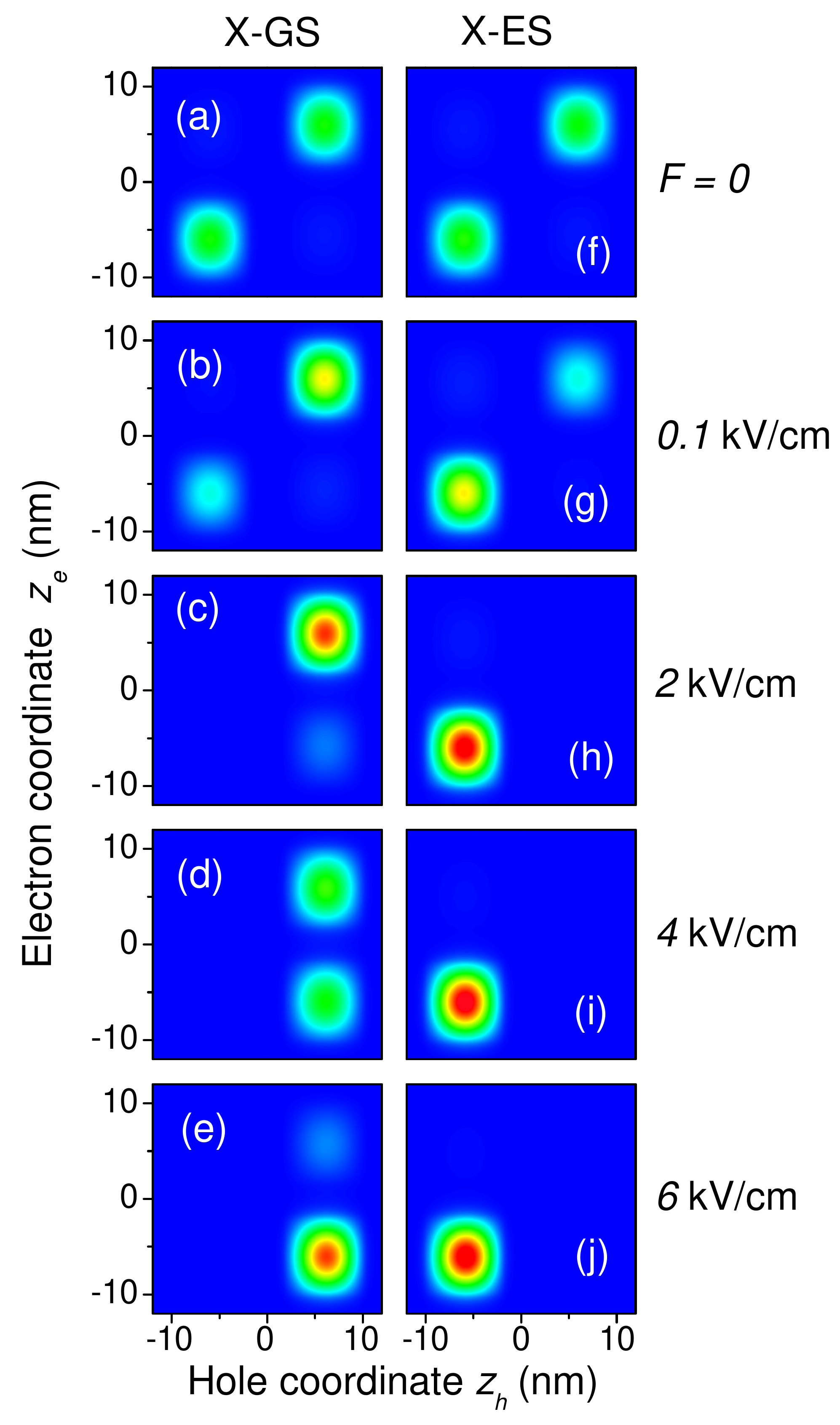}
 \caption{Probability distributions $\int^{\infty}_{0} \left|\Phi(z_{e},z_{h},\rho)\right|^2 2 \pi \rho d \rho$ calculated for the exciton ground state X-GS (a-e) and excited state X-ES (f-j), for different values of the electric field.
 \label{fig:psi2d}}
  \end{center}
  \end{figure}

The Coulomb coupling matrix elements $V_{13}\approx V_{24}$ are responsible for the mixing of direct and indirect pair states and in particular for a {\em crossover} of the ground exciton state from direct to indirect type as the EF grows.
Owing to this off-diagonal coupling, the Stark red-shifted indirect state $e1h1$ ($n=1$) has a remarkable anticrossing with the direct pair state $e2h1$ ($n=3$), weakly dependent on the EF. This anticrossing takes place at about $F=5$\,kV/cm and is clearly seen in Fig.\,\ref{fig:osc}. Indeed, the exciton ground state (X-GS) experiences a crossover from direct to indirect exciton: The oscillator strength of the X-GS has its maximum at $F=0$ and then drops quickly with increasing EF as seen in Fig.\,\ref{fig:osc}.
Since all other matrix elements including $V_{12}$ are generally much smaller, the other direct pair state $e1h2$ ($n=2$) remains unaffected and is only Coulomb shifted by the diagonal element $V_{22}$. Although it is strongly coupled via $V_{24}$ to the other indirect state $e2h2$ ($n=4$), the latter is Stark blue-shifted and thus significantly detuned from $e1h2$ having a minor effect on it. As a result, the energy position of the brightest exciton excited state (X-ES), which has the maximum oscillator strength in the excitonic spectrum, remains practically unchanged.

The excitonic states shown in Fig.\,4(a) and (b) are calculated with different in-plane  exciton confinement radius,
$R=200$\,nm  and 800\,nm, respectively [see Eq.\,(\ref{circle})]. In the latter case the excitonic continuum has a much finer discretization that makes more clear which states belong to the continuum and which are the true bound states having more or less isolated positions, weakly dependent on $R$. For example, the $2S$ and $3S$ states of the  indirect exciton are clearly identified in Fig.\,4(b): They lie just below the discretized continuum onset  and are down-shifted with $F$ almost parallel to the exciton ground state.
Higher excited states of the direct exciton are also well seen in Fig.\,4: They are deep in the continuum (7-8\,meV above the brightest direct state) and are weakly dependent on the EF.

 \begin{figure}[t]
 \begin{center}
\includegraphics[width=\columnwidth]{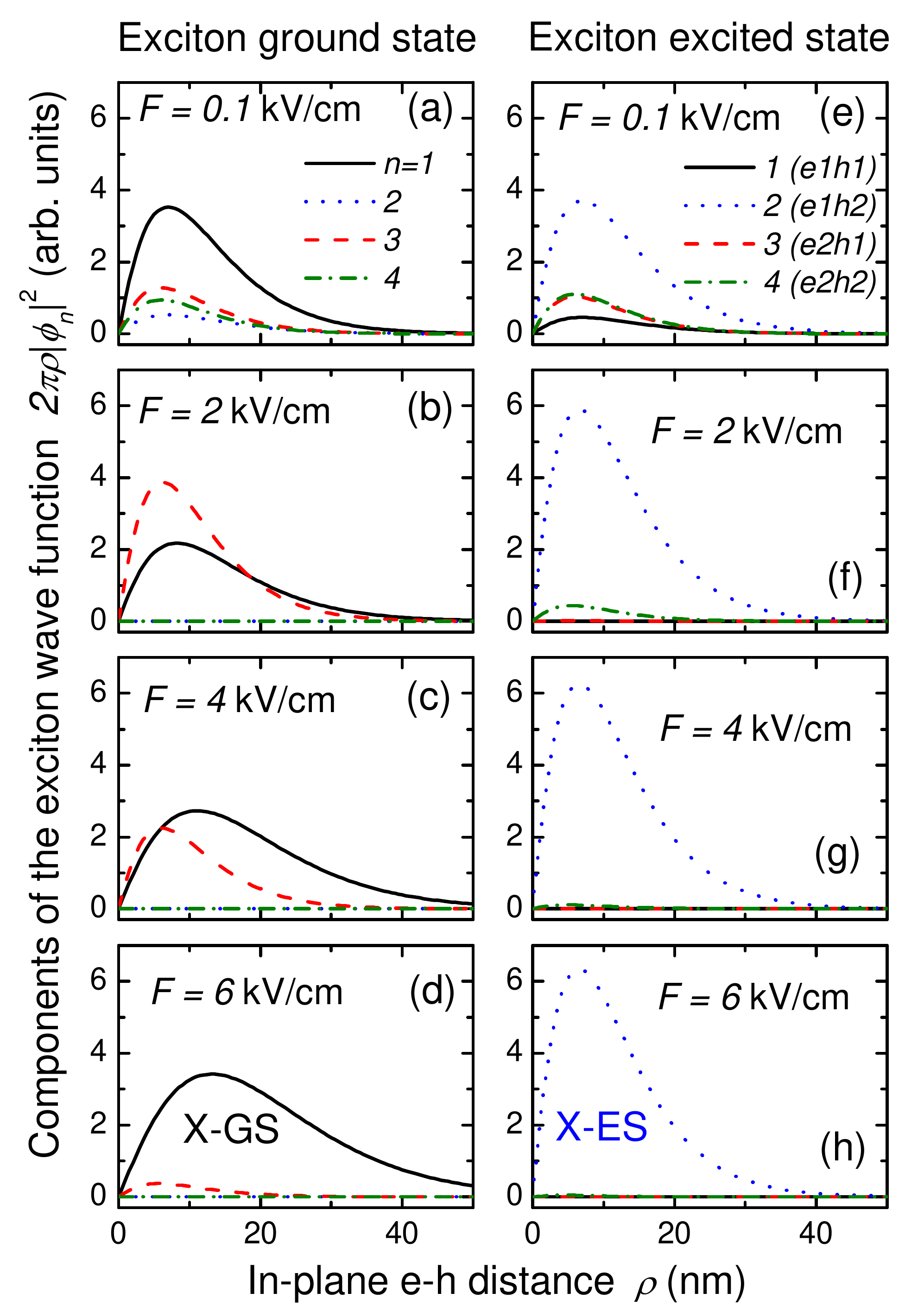}
 \caption{Radial components $2\pi\rho|\phi_n(\rho)|^2$ of the exciton wave function calculated for the ground state X-GS (a-d) and excited state X-ES (e-h), for different values of the electric field. \label{fig:psirho}}
  \end{center}
  \end{figure}

 \begin{figure}[t]
 \begin{center}
\includegraphics[width=\columnwidth]{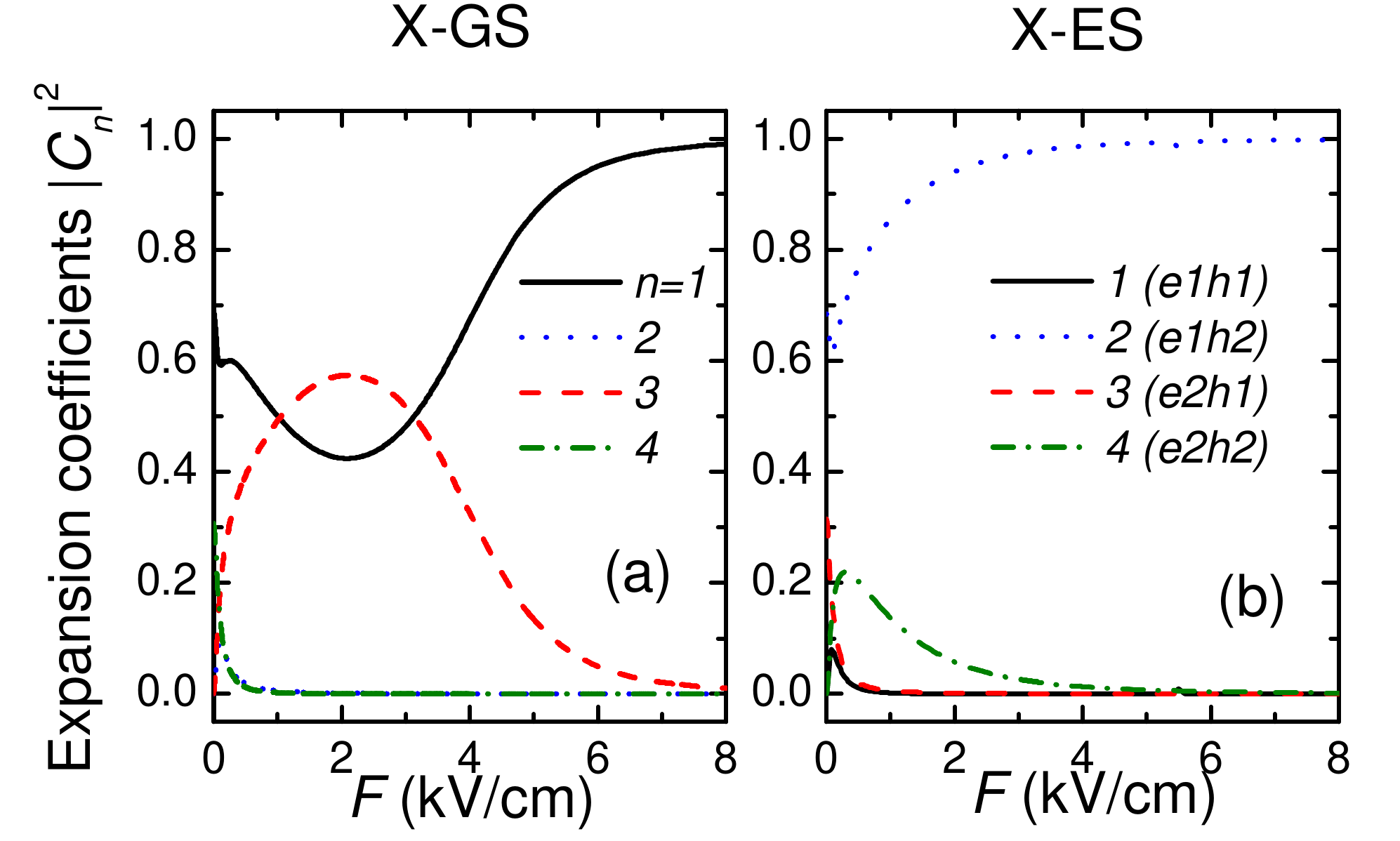}
 \caption{ Coefficients $C_n$ of the expansion of the exciton wave function into e-h pair states, calculated for the ground state X-GS (a) and excited state X-ES (b) as functions of the electric field $F$.
 \label{fig:Cn}}
  \end{center}
  \end{figure}

  \begin{figure}[t]
  \begin{center}
\includegraphics[width=70mm]{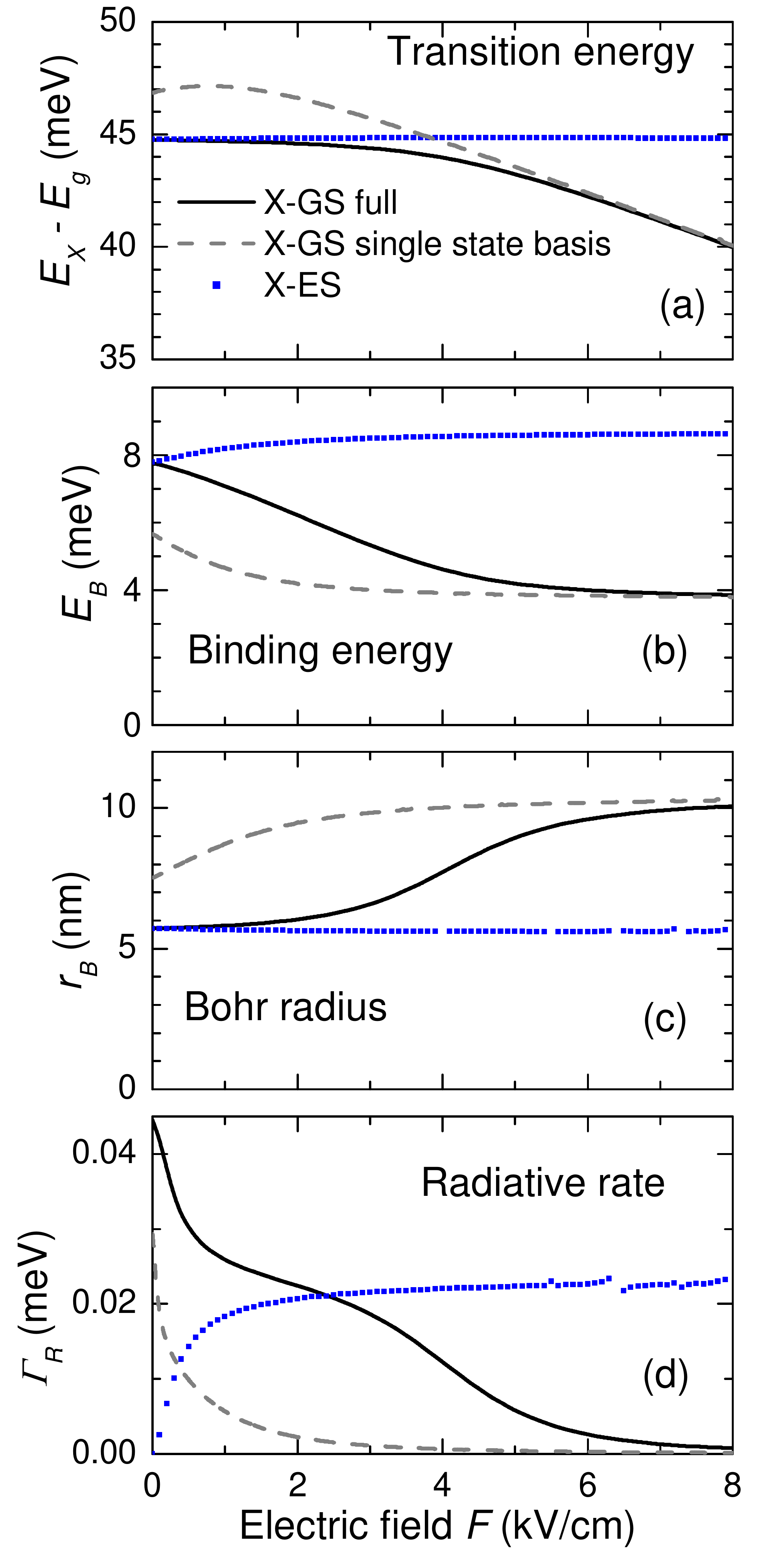}
  \caption{(a) Optical transition energy $E_X$, (b) binding energy $E_b$, (c) in-plane Bohr radius  $r_{_B}=\sqrt{\langle \rho^2\rangle}$, and (d) radiative rate $\Gamma_R$ of the exciton ground state X-GS (solid lines) and excited state X-ES (full squares)
  as functions of the electric field $F$. Dashed lines are the single state basis calculation of the X-GS.
\label{fig:Eb}}
  \end{center}
  \end{figure}

Let us consider the properties of the X-GS and X-ES, and in particular the direct-to-indirect (D-I) crossover
in more detail. The contour plots in Fig.\,\ref{fig:psi2d} show localization of the X-GS and X-ES across the CQW structure, for different values of $F$. At zero EF, owing to the symmetry of the system, both states have two identical maxima on the main diagonal $z_e=z_h$ [Fig.\,\ref{fig:psi2d}(a) and (f)], which refer to the direct nature of both excitonic states. With increasing EF one of the two peaks becomes smaller and then vanishes, see Fig.\,\ref{fig:psi2d}(b,c) and (g,h). The states become asymmetric
having both carriers localized either in the right QW (in X-GS) or in the left QW (in X-ES). Further increase of the EF up to $F=6$\,kV/cm leads to the X-GS switching from direct to indirect state: The peak moves away from the main diagonal towards the bottom right corner, see Fig.\,\ref{fig:psi2d}(c)-(e). This result is in good agreement with previous theoretical findings\cite{Arapan05,Szymanska03} and experimental observations.\cite{Butov99} At the same time the X-ES remains unchanged. The dominant hole component in the X-ES changes from $h2$ to $h3$ and from $h3$ to $h4$ when the system passes through the anticrossings of the hole levels at $F=27.8$\,kV/cm and 74\,kV/cm, respectively. The X-ES remains the brightest state in the excitonic spectrum having at the same time a very weak dependence on the EF that emphasizes its direct nature.

The D-I crossover of the X-GS is demonstrated also in Fig.\,6 where different radial components $\phi_n(\rho)$ of the X-GS and X-ES WFs are plotted. At $F=2$\,kV/cm the direct e-h pair state $n=3$ has the dominant contribution to the X-GS. This state is strongly coupled to the indirect pair state $n=1$ via the Coulomb matrix element $V_{13}$, as discussed above. As a result of this anticrossing, $\phi_1$ grows and $\phi_3$ reduces with the EF.
Nothing similar happens to the X-ES. The latter is always dominated by the $n=2$ pair state which is coupled to the significantly detuned $n=4$ state. Therefore, only a minor contribution of the $\phi_4$ to the X-ES can be seen in Fig.\,6(f)-(h).
The picture is completely different for smaller EFs. At $F=0$ the symmetric $n=1$ pair state is Coulomb coupled to the other symmetric state ($n=4$) producing bright X-GS and dark X-ES. These symmetric pair states do not interact with the antisymmetric states ($n=2$ and 3) which are, in turn,
coupled to each other. The transition from such symmetric coupling to the above considered D-I coupling
has a rather narrow interval of small values of the EF and involves interaction of all four e-h pair states. In fact, for $F=0.1$\,kV/cm one can see in Fig.\,6(a) that all four components have comparable contributions to
the X-GS and X-ES WFs. Figure~7 summarizes our analysis showing the field dependence of the e-h pair amplitudes $C_n$ introduced in Eq.\,(\ref{C-intro}). It demonstrates the prominent D-I crossover in the X-GS, a much weaker D-I coupling in the X-ES, and a very quick transition from symmetric coupling to D-I coupling, seen in the WFs of both X-GS and X-ES.

Figure~8 shows the field dependence of the optical transition energies $E_X$ and binding energies $E_b$ of X-GS and X-ES, as well as their in-plane exciton Bohr radii $r_{_B}=\sqrt{\langle \rho^2\rangle}$ and the radiative linewidths $\Gamma_R$. The latter are  calculated via Eq.\,(\ref{Rad}) assuming $d_{cv}=0.6$\,nm.\cite{dcv,dcv2}  While all these parameters for the X-GS change dramatically when the EF increases from 2\,kV/cm to 8\,kV/cm, the X-ES remains practically unaffected. The X-GS energy exhibits a considerable Stark shift, for $F>6$\,kV/cm almost linear in $F$ [Fig.\,\ref{fig:Eb}(a)], due to the electron-hole separation typical for the indirect exciton. The X-ES in turn has a very weak field dependence due to a much smaller dipole moment of the direct exciton, but in a larger range of the EF values, the X-ES transition energy is also red-shifted (by 12\,meV at $F=100$\,kV/cm), as a result of an EF-induced electron-hole separation within the same QW.

The X-GS binding energy, $E_b=E_1^{(0)}-E_X$, drops from 8\,meV down to 4\,meV [Fig.\,\ref{fig:Eb}(b)], as a result of the transition from direct to indirect Coulomb coupling. For the X-ES the binding energy is defined as the energy distance from X-ES to its own continuum onset: $E_b=E_2^{(0)}-E_X$. This is done because the lowest-energy pair state, $e1h1$, has negligible (for $F>1$\,kV/cm) contribution to the X-ES, and thus this exciton state remains bound even though its energy $E_X$ is higher than the first continuum onset $E_1^{(0)}$. The X-ES binding energy is almost unchanged, as the EF has no effect on the direct nature of this state. The Bohr radius [Fig.\,\ref{fig:Eb}(c)] is fully correlated with the binding energy, increasing with the EF almost by a factor of 2 for the X-GS and showing no change for the X-ES.

The D-I crossover is accompanied by a dramatic decrease of the radiative linewidth of the X-GS
(proportional to its oscillator strength), see Fig.\,\ref{fig:Eb}(d).
This happens because the EF makes the electron-hole separation larger reducing
the overlap of the electron and hole WFs.
The X-ES, in turn, being strictly dark at $F=0$, becomes bright in a finite EF, and its linewidth
quickly increases with the EF up to the half of the maximum linewidth of the X-GS.
Further increase of the EF does not change the X-ES radiative rate much but $\Gamma_R$ experiences
some influence of higher ESs, see fluctuations in Fig.\,\ref{fig:Eb}(d).
  \begin{figure}[t]
  \begin{center}
\includegraphics[width=\columnwidth]{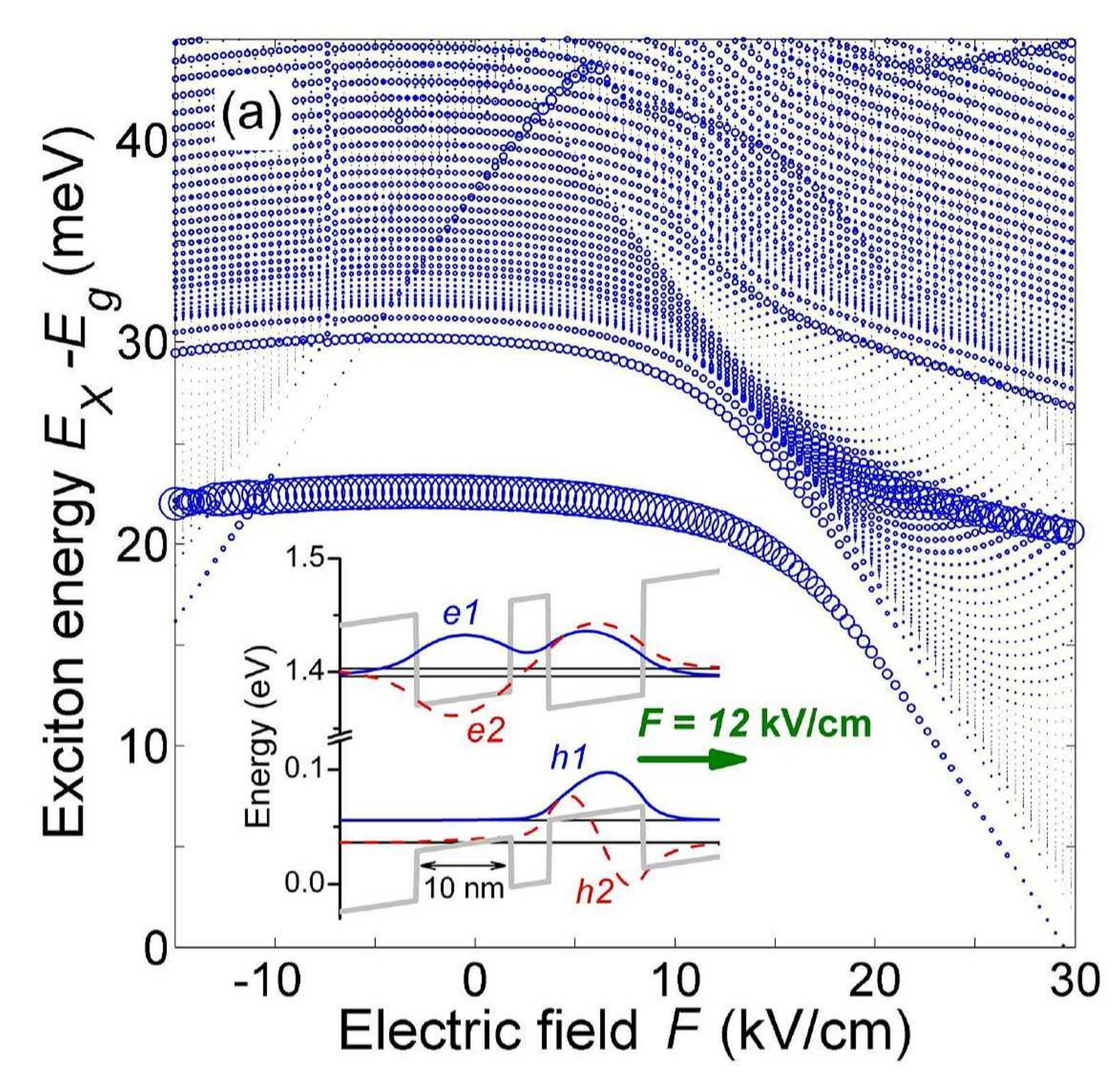}
\includegraphics[width=\columnwidth]{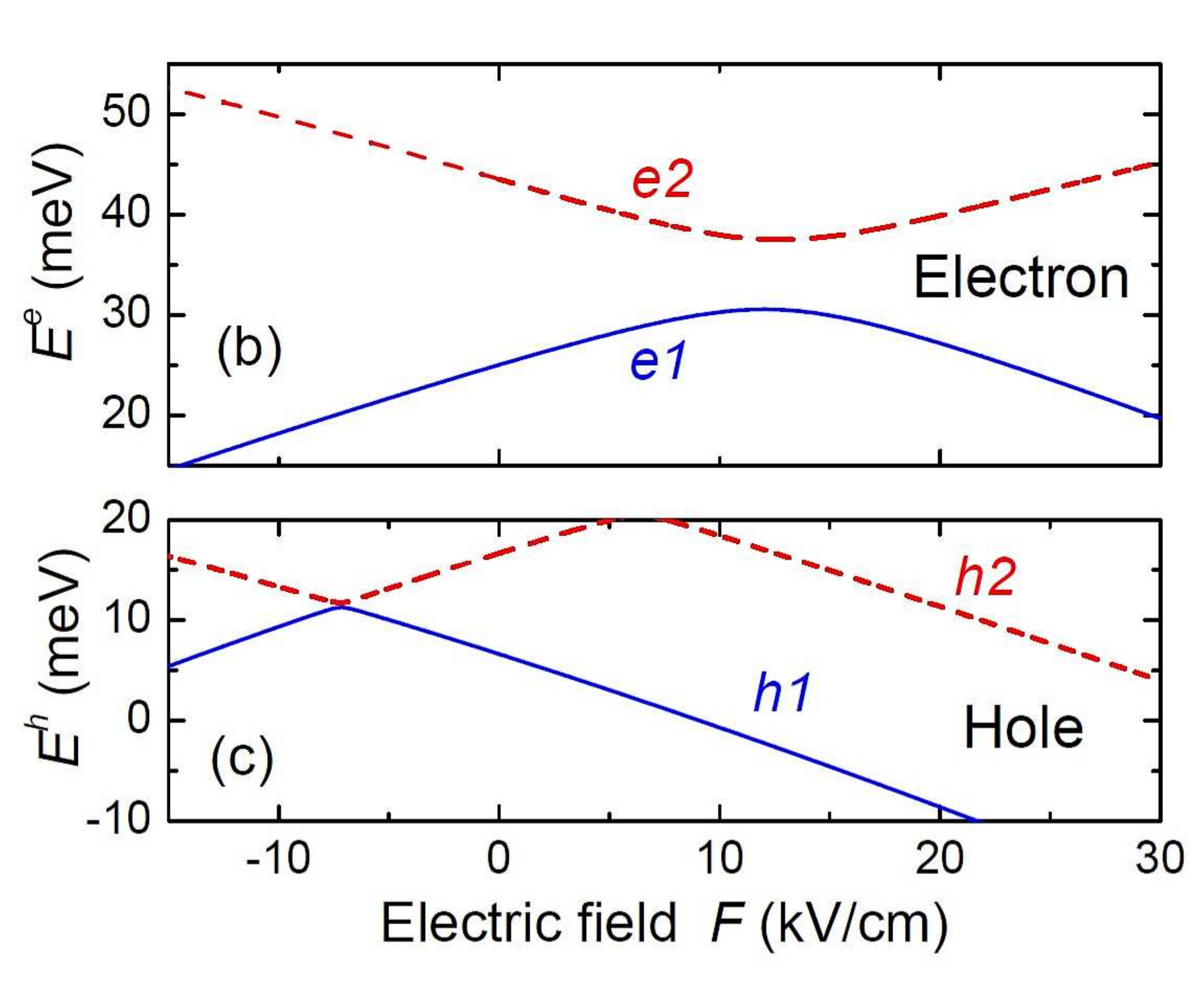}
  \caption{ (a) Exciton energies and oscillator strengths (circle area) and (b,c) electron and hole subbands in
asymmetric In$_{0.08}$Ga$_{0.92}$As/GaAs/In$_{0.1}$Ga$_{0.9}$As (10\,nm/4\,nm/10\,nm) CQW structure as functions of the applied electric field $F$. The inset shows CQW band structure and direction of the electric field, as well as the electron and hole wave functions and energy levels of the ground and first excited quantized states, for $F=12$\,kV/cm.
\label{fig:new}}
  \end{center}
  \end{figure}

To reproduce some previous simulations\cite{Galbraith89} and to compare them with our full calculation,
we have restricted our basis to the electron and hole GSs only, neglecting
any ESs quantized in the growth direction.
In other words, we have taken into account only one e-h pair state $e1h1$,
choosing $N=1$ in the expansion Eq.\,(\ref{ex-wf}).
This single-state basis (SSB) calculation shows considerably different results compared to the
full calculation in all four plots in Fig.\,\ref{fig:Eb}, see dashed curves.
The reason for such difference is obvious: Figure~\ref{fig:Cn}(a) clearly demonstrates the importance
of taking into account higher e-h pair states for proper description of
the D-I crossover of the X-GS, and in particular the role in such crossover of state $e2h1$ which is missing in the SSB calculation. This state is a direct pair state
(at least for $F>0.5$\,kV/cm), so that neglecting it in the X-GS calculation, as done e.g. in Ref.\,\onlinecite{Galbraith89},
underestimates the exciton binding energy by a factor of 1.5 and the exciton radiative linewidth by an order of magnitude.  Nevertheless, at larger EFs ($F>10$\,kV/cm) the SSB model adequately
describes the properties of the X-GS as such an indirect exciton state is strongly dominated by $e1h1$.
  \begin{figure}[t]
  \begin{center}
\includegraphics[width=\columnwidth]{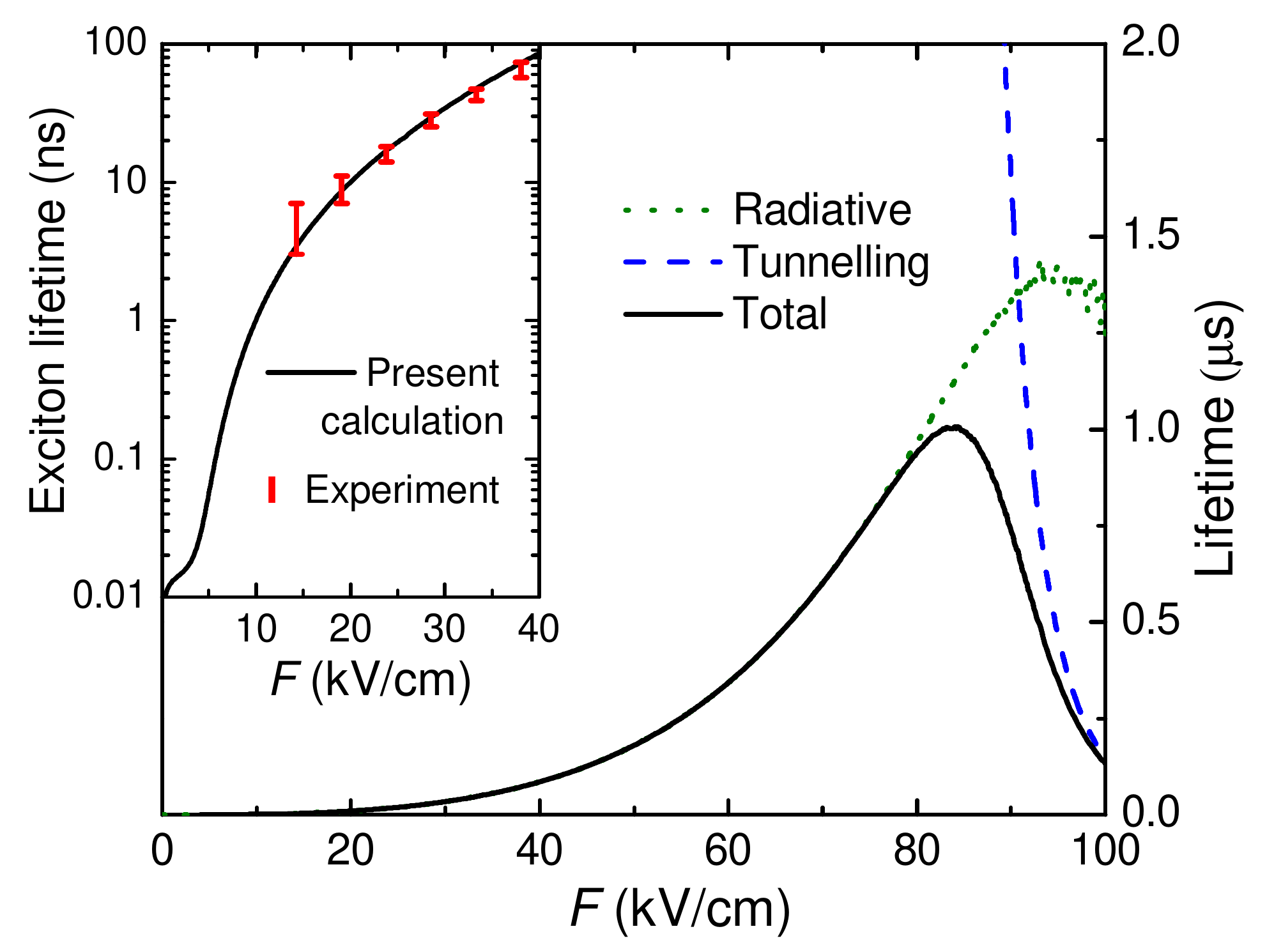}
  \caption{ Radiative (dotted curve), tunneling (dashed curve), and total lifetimes (solid curve) of the exciton ground state X-GS as functions of the applied electric field. Inset: logarithmic plot of the X-GS total lifetime in comparison with measured photoluminescence decay times (error bars) extracted from Ref.~\onlinecite{Butov99}.
\label{fig:tauR}}
  \end{center}
  \end{figure}

Finally, Fig.\,9 shows our simulation for an asymmetric
In$_{0.08}$Ga$_{0.92}$As/GaAs/In$_{0.1}$Ga$_{0.9}$As (10\,nm/4\,nm/10\,nm) CQW structure
used in Ref.~\onlinecite{Christmann11}. As expected, the energy spectra are asymmetric with respect to the EF direction. In the right part of the exciton energy spectrum [Fig.\,9(a)],  at around 16\,kV/cm, a Coulomb-induced anticrossing is observed, which is similar to that seen in Fig.\,4 at $F=5$\,kV/cm. The physical mechanism which causes this anticrossing is essentially the same as in the case of symmetric CQWs, but the anticrossing takes place at much higher values of the electric field. This is because at around $F=12$\,kV/cm (4-5\,kV/cm below the anticrossing) the applied EF almost compensates the asymmetry in the conduction band structure, and thus at this value of $F$ the properties of the asymmetric CQW can resemble those of the symmetric CQW at $F=0$. In fact, a repulsion of the electron subbands is seen in Fig.\,9(b) as well as a formation of symmetric and antisymmetric electronic states, as is clear from the inset in Fig.\,9(a).

\subsection{Indirect exciton lifetime}

A CQW exciton can escape from the system using the following two major channels: It can either recombine
by emitting a photon or the electron and/or hole can tunnel through the external barrier with the help of the EF.
We concentrate here on the X-GS only, and combining both channels together, the total exciton lifetime takes the form
\begin{equation}
\frac{1}{\tau} = \frac{1}{\tau_{_{R}}} + \frac{1}{\tau_{_{T}}}\,,
\end{equation}
where $\tau_{_{R}} = \hbar/(2 \Gamma_{R})$ is the exciton radiative lifetime.
As for the tunneling time $\tau_{_{T}}$, we take into account the lowest pair state only.
This is a valid approximation because when other pair states contribute to the X-GS and thus can have
some effect on the exciton tunneling, its lifetime is strongly dominated by the radiative channel as can be seen
in Fig.\,\ref{fig:tauR}. In particular, the tunneling lifetime $\tau_{_{T}} = \hbar/(\Gamma^{e}_1+\Gamma^{h}_1)$ is much longer than the radiative one up to $F=80$\,kV/cm.
The D-I crossover of the X-GS is  accompanied by a monotonous growth of its radiative lifetime, in accordance with Fig.\,8(d). Indeed, a direct exciton has a short lifetime because the carriers are in the same well,
so that they can easily recombine. Increasing the electron-hole separation leads to a dramatic increase of the radiative lifetime.
The probability of the electron and hole tunneling also increases with EF. At some point the tunneling time becomes comparable to the radiative lifetime and then starts to dominate.
We have also compared the calculated radiative lifetime for the X-GS with the experimental
results taken from Ref.\,\onlinecite{Butov99} where the low-temperature excitonic photoluminescence
was measured in 8--4--8\,nm GaAs/Al$_{0.33}$Ga$_{0.67}$As CQWs.
The inset to Fig.\,\ref{fig:tauR} demonstrates a quantitative agreement between
the experiment and the present theory.

\subsection{Absorption spectrum}

We also calculate the exciton absorption coefficient in a 8--4--8\,nm GaAs/Al$_{0.33}$Ga$_{0.67}$As CQW at different frequencies of the incoming light. Using the Lorentzian model of absorbing oscillators\cite{Elliot57,Andreani95} and leaving out a common prefactor, the  absorption of the exciton with zero in-plane momentum takes the form
\begin{equation}
\alpha(\omega) = \sum_{\nu}\Gamma_{R,\nu}\, \frac{\Gamma_{R,\nu}}{(\hbar\omega-E_{\nu})^{2}+\Gamma_{R,\nu}^{2}}\,,
 \label{Abs}
\end{equation}
where the index $\nu$ labels all possible excitonic states calculated in the theory, and $E_\nu$ and $\Gamma_{R,\nu}$
stand for their energies and radiative linewidths.

  \begin{figure}[t]
  \begin{center}
\includegraphics[width=\columnwidth]{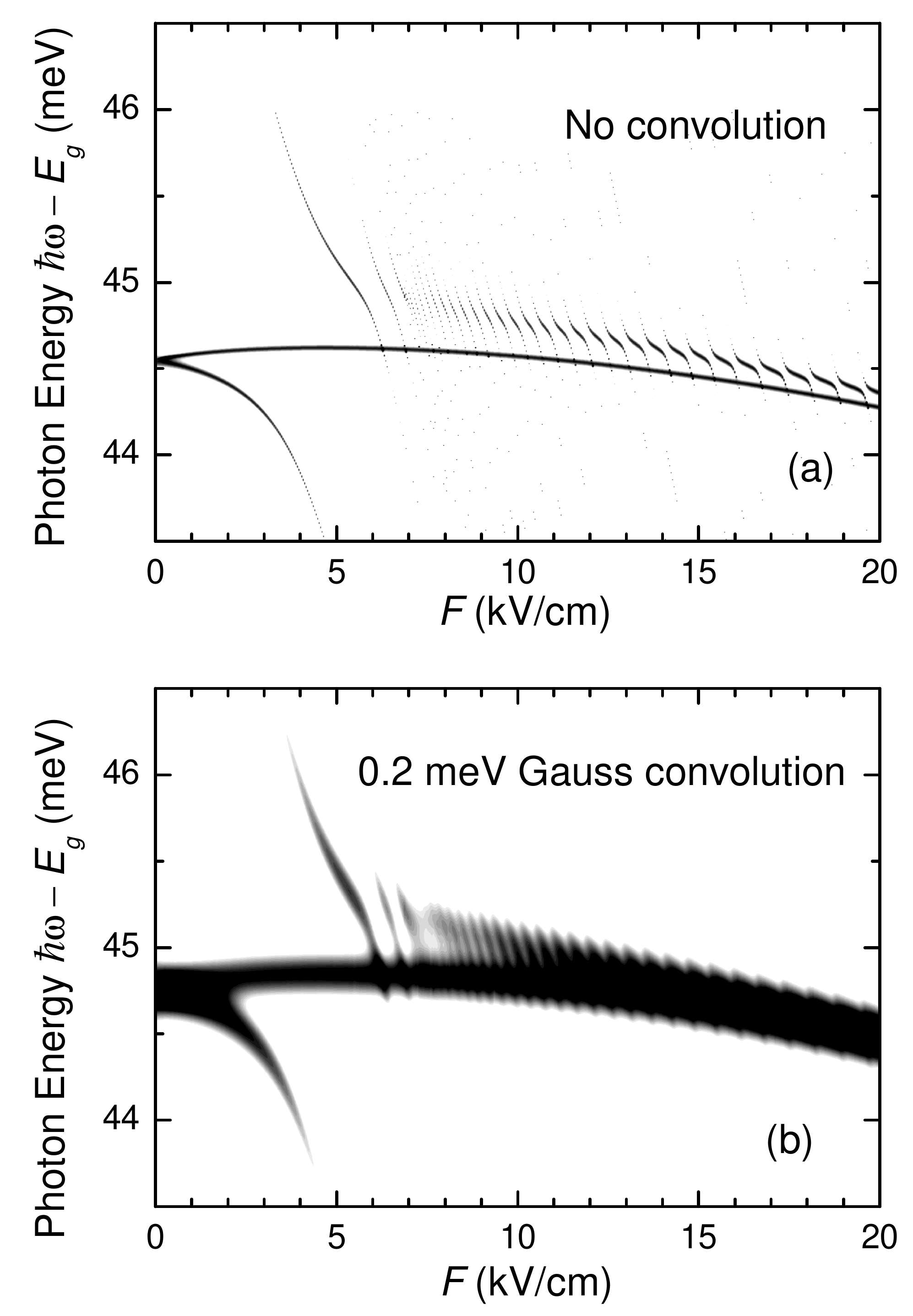}
  \caption{(a) Electric field dependence of the full excitonic absorption spectrum. (b) The same spectra convoluted
  with a Gaussian function with 0.2\,meV full width at half maximum.
 \label{fig:AbsG}}
  \end{center}
\end{figure}

Figure~\ref{fig:AbsG}(a) shows the absorption spectrum calculated using Eq.\,(\ref{Abs}) and
in-plane  exciton confinement radius $R=800$\,nm. All lines in the absorption
have very narrow radiative widths ($<0.1$\,meV) which are calculated according to Eqs.(\ref{Osc}) and (\ref{Rad}).
The Lorentzian model however has an obvious artefact: Although the spectrum properly reproduces the linewidths,
all lines have the same peak height, and the fact that not all of them are seen in Fig.\,\ref{fig:AbsG}(a)
is only due to the resolution of the plot. To improve on this and also to take into account the effect of inhomogeneous
line broadening in realistic CQW structures, we make a Gauss convolution of the spectrum:
$A(\omega)= \int_{-\infty}^\infty\alpha(\omega^{'}) g(\omega-\omega^{'})d \omega^{'}$ with a
normalized Gauss function  $g(\omega)=(\Delta\sqrt{\pi})^{-1}e^{-\omega^{2}/\Delta^{2}}$.
The convoluted spectrum with the full width at half maximum $2\sqrt{\ln 2}\Delta =0.2$\,meV is shown
in Fig.\,\ref{fig:AbsG}(b). All lines now have
almost the same width but their peak maxima now reflect the optical
strength of the corresponding exciton states.

The two lowest excitonic states, X-GS and X-ES, are well resolved in the spectrum up to $F=5$\,kV/cm.
Then the indirect exciton  X-GS loses its optical activity. The bright direct X-ES line superimposes with
higher ESs and discretized continuum of the indirect exciton, all line merging up together at higher EFs.

\subsection{Numerical code}

We have provided a freely available on-line software\cite{OnlineCode} which calculates
the optical properties of the direct and indirect excitons in a CQW
structure in a perpendicular EF.
The software has a user-friendly interface and produces
an on-screen output of the requested calculation as well as a pdf version of
the same data and plots.
For the input, the user has an option either to choose a symmetric AlGaAs CQW
inserting required structural parameters ($L_b$, $L_w$, and $x$),
or to take other semiconductor CQW by inserting its structural and
material parameters. The value of the EF is also needed for the input.
In this numerical code, we use four basis states for the exciton WF, taking into account only the ground and the first excited states for the electron and hole. The exciton confinement radius is set to $R=200$\,nm. The transition energies and the oscillator strengths of exciton excited states are also included in the output file. The absorption spectrum is calculated with 0.1\,meV FWHM Gaussian convolution.

\section{Conclusions}

We have studied the effect the electric field has on excitonic states in
AlGaAs and InGaAs CQWs. To do this we have developed an
efficient numerical approach which is based on expanding the exciton wave function
into uncorrelated electron-hole
pair states and solving in the real space a matrix Schr\"odinger equation for the CQW exciton.
Using this approach  we have calculated the energies and the
wave functions of exciton  states in the CQW and studied their optical properties in the presence of
electric field, by addressing such important parameters of the exciton as its binding energy,
Bohr radius, radiative and tunneling times, and optical absorption spectrum.
While we are able to  calculate a large number of exciton states, we have concentrated on two
most important ones, the exciton ground state X-GS and the brightest excited state X-ES. We have shown that
the Coulomb coupling between direct and indirect pair states leads to a prominent effect in the presence of the electric field: a direct-to-indirect crossover of the X-GS. At the same time, the properties of the X-ES remain almost unchanged. Finally, we have calculated the exciton lifetime which consists of two main components, radiative and tunneling times, and shown that the latter reduces the total lifetime at higher electric fields.

\acknowledgments

We thank Jeremy Baumberg, Wolfgang Langbein, Roland Zimmermann for useful discussions, and Richard Frewin for technical support of the website.
K.\,S. acknowledges support of the Royal Thai Government. L.\,M. acknowledges FP7-REGPOT-2008-1, Project BIOSOLENUTI No 229927
for financial support.

\end{document}